\documentclass[12pt,letterpaper]{JHEP3}
\usepackage{epsfig}
\usepackage{caption}
\usepackage{graphicx}
\usepackage[english]{babel}
\usepackage{amsmath}

\usepackage{cite}

\graphicspath{{figures/}}

\title{ Secret Loss of Unitarity due to the Classical Background }

\author{I-Sheng Yang \\
Canadian Institute of Theoretical Astrophysics, \\
60 St George St, Toronto, ON M5S 3H8, Canada \\ 
and \\
Perimeter Institute of Theoretical Physics, \\
31 Caroline Street North, Waterloo, ON N2L 2Y5, Canada
}

\abstract{
We show that a quantum subsystem can become significantly entangled with a classical background through a process with little or none semi-classical back-reactions.
We study two quantum harmonic oscillators coupled to each other in a time-independent Hamiltonian. 
We compare it to its semi-classical approximation in which one of the oscillators is treated as the classical background.
In this approximation, the remaining quantum oscillator has an effective Hamiltonian which is time-dependent, and its evolution appears to be unitary.
However, in the fully quantum model, the two oscillators can entangle with each other.
Thus the unitarity of either individual oscillator is never guaranteed.
We derive the critical time scale after which the unitarity of either individual oscillator is irrevocably lost.
In particular, we give an example that in the adiabatic limit, unitarity is lost before other relevant questions can be addressed.
}

\begin{document}

\section{Introduction and Outline}

Unitarity is one of the defining properties of quantum mechanics and quantum field theory.
The ideal concept of unitarity should be applied to the {\it whole} system, namely the entire universe.
But in practice, no one ever tries to solve the quantum mechanics of the whole system.
We always focus on small subsystems, and pretend that the rest of the world forms a {\it classical background}.
In other words, whenever we write down a practical quantum mechanical description, strictly speaking, it is always a {\it semi-classical approximation}.
From the whole system, we artificially choose a subsystem to describe by quantum mechanics, while treating the rest of the system as being classical.

Of course, such semi-classical approximation is inevitable for many practical reasons, and we are not advocating alternatives.
But since it is an approximation, it must come with a validity condition, something that warns us when it breaks down.
A common approach is to calculate semi-classical back-reactions.
There are many conservation laws which both classical and quantum physics have to obey, such as energy and momentum.
In a semi-classical model, we can demand that the quantum expectation value (from the quantum subsystem) and classical value (from the classical background) of the conserved charges are together conserved.
These ``quantum + classical'' conservation laws allow us to calculate the back-reaction of the quantum subsystem on the classical background.
Sometimes we reasoned that if the back-reactions are not significant, the semi-classical approximation is reliable.
\footnote{For example, one calculates Hawking radiation in a fixed geometry, and then use conservation of energy to argue that the black hole has to evaporate \cite{Haw74}.
Since the evaporation is so slow, it is a small correction to the fixed geometry.
Sometimes we take this as the reason to trust such semi-classical approximation.}

The smallness of such back-reactions can be a necessary condition for the validity of semi-classical models, but we should not expect it to be sufficient.
The apparently unitary evolution of a quantum state usually contains infinitely more information than all the conserved charges.
It is natural to conjecture that only a much more stringent condition can grant us the right to trust the full details of the quantum evolution in a semi-classical description.
In this paper, we will give an example to support such conjecture, and also explicitly derive the condition for which the unitary quantum evolution in the semi-classical description can be trusted.

We are inspired by Unruh's example of ``decoherence without dissipation'' \cite{Unr12}. 
He had a toy-model in which two quantum subsystems can become entangled without transferring energy to each other.
We will show that similar things can happen when one of the quantum subsystems is well-approximated by a classical system.
This leads to a semi-classical approximation with little back-reactions, yet the apparently unitary evolution of the remaining quantum subsystem is an illusion.

Our particular model is about two coupled harmonic oscillators.
We focus on the situation that their frequencies are far from resonant, the second oscillator starts in a coherent state, and its energy is much larger than the first oscillator.
In this situation, the expected position of the second oscillator will remain close to what we expect from a coherent state for a long time. 
Thus it is very natural to treat the it as being classical.
In such approximation, the first oscillator will have a time-dependent effective frequency, but its evolution appears to be unitary forever.

In this paper, we will show that the two oscillators can become significantly entangled, while classical back-reactions are negligible.
In other words, the evolution of the first oscillator will decohere, but within the semi-classical approximation there is no way to raise a flag.
The apparent unitarity in the semi-classical approximation is secretly lost.


We derive the time scale at which this unitarity loss happens.
Such time scale can be independently tuned from the parameters in the semi-classical approximation.
For example, in the adiabatic limit of the semi-classical model, it takes a very long time for the effective frequency for the first oscillator to change.
Most of the interesting physical questions are about what happens after the frequency change. We can easily arrange that unitarity is lost way before those interesting physical questions can be addressed.

We should emphasize that such unitarity loss does not invalidate the entire semi-classical approximation. 
For example, in the adiabatic limit, we can describe the first oscillator by the eigenstates of its instantaneous effective frequency.
The occupation numbers in this basis stay the same, and that can still be true even if unitarity is lost.
The only difference is that instead of a unitary transformation,
$(|0\rangle_{w_i} + |1\rangle_{w_i}) \rightarrow (|0\rangle_{w_f} + |1\rangle_{w_f})$,
a pure state density matrix evolving into a mixed state,
$(|0\rangle_{w_i} + |1\rangle_{w_i}) (\langle0|_{w_i} + \langle1|_{w_i}) 
\rightarrow (|0\rangle_{w_f}\langle0|_{w_f} + |1\rangle_{w_f}\langle1|_{w_f})$.
Thus, we are not questioning the apparent success in many applications of the semi-classical approximations.
We simply point out that the apparent unitarity in semi-classical approximations might need further scrutinization. 

If we believe in the existence of quantum gravity, then quantum field theory in curved space time (QFT-CST) is essentially a semi-classical approximation of that.
A few interesting calculations in QFT-CST, such as quantum fluctuations during inflation \cite{LidLyt93} and Hawking radiation from black holes \cite{Haw74}, closely resemble our simple model of a harmonic oscillator with a time-dependent frequency.
Thus our work not only points out the potential unitarity issue in QFT-CST, but it also provides a first step to address such problem.

QFT-CST is considered quite successful by many physicists.
The density perturbation of the universe is considered an observed proof to QFT-CST during inflation, and the temperature/spectrum of Hawking radiation is an essential component of black hole thermodynamics.
We wish to emphasize again that questioning the unitarity of QFT-CST is not in conflict with those widely accepted successes.
Those results of QFT-CST are analogous to predicting the occupation number in our toy model.
It is possible to lose unitarity without changing those predictions.

For inflationary perturbations, the unitarity of QFT-CST can in-principle be checked by cosmological Bell inequalities in some inflation models \cite{Mal15}.
But even for those models, the corresponding observation is still out of our reach at this moment.
For Hawking radiation, the unitarity of QFT-CST actually leads to the famous information paradox
\cite{Haw76a,AMPS}.
The opportunity to deny the apparent unitarity without changing the thermal properties of Hawking radiation provides an elegant way to resolve the paradox, as discussed in \cite{OsuPag16,BakKod17}.

The rest of the paper goes like the following.
In Sec.\ref{sec-analytic}, we write down the exact Hamiltonian and initial states for both the semi-classical model and the fully quantum model.
In the fully quantum model, we use non-degenerate perturbation theory to derive how the two oscillators can become entangled with each other.
That leads to Eq.~(\ref{eq-entangle}), which is the time scale of unitarity loss and the main result of this paper.
Technically speaking, non-degenerate perturbation theory is only marginally applicable in our case.
Degenerate perturbation theory is required to get the exact answer.
We give a physical argument for why the answer we get from non-degenerate perturbation theory is good enough.
In Sec.\ref{sec-numeric}, we use degenerate perturbation theory and numerically solve two examples.
This conforms our analytical derivation, and also reveals another interesting effect that we provide an analytical explanation in Sec.\ref{sec-Losses}.
In Sec.\ref{sec-dis}, we summarize our results and and discuss future directions.

\section{Analytical Approach}
\label{sec-analytic}

\subsection{The semi-classical model}

Consider a quantum harmonic oscillator ``$A$'' with a time-dependent natural frequency.
\begin{eqnarray}
H_A &=& \frac{P_A^2}{2} + 
\left(\frac{w_0^2+w_A^2}{4} + \frac{w_0^2-w_A^2}{4}\cos w_Bt \right)X_A^2~.
\label{eq-HA}
\end{eqnarray}
Basically, the effective frequency starts as $w_0$ at $t=0$, smoothly changes to $w_A$ at $t=\pi w_B^{-1}$, and continues to oscillate between $w_0$ and $w_A$ in this manner.
Similar time-dependent Hamiltonian has been widely studied in quantum mechanics and quantum field theory for its interesting properties.
For example, there can be particle productions.
Starting in the ground state of $w_0$, the system can be excited at a later time even if the effective frequency returns to the same value.
\begin{equation}
|\phi_A(2\pi w_B^{-1})\rangle = e^{-i\int_0^{2\pi w_B^{-1}} H_A dt}|0\rangle_{w_0} = 
\sum_n c_n|n\rangle_{w_0} ~,
\label{eq-unitary}
\end{equation}
where $c_{n>0}\neq0$ in general.

The main purpose of this paper is not about those interesting properties.
We are interested in a hidden assumption that is implicit when we wrote down Eq.~(\ref{eq-unitary}).
This equation clearly shows that in this model, if we start in a pure state, it will remain to be pure {\it forever}.
Namely, the evolution of this harmonic oscillator alone is unitary.
Such assumption should not be taken for granted.
We will provide a concrete example to show how it breaks down.

\subsection{The quantum model}

First of all, the apparently time-dependent Hamiltonian can be traced back to the evolution of some ``classical background''.
The particular form of Eq.~(\ref{eq-HA}) is chosen such that the time-dependent classical background is another harmonic oscillator ``$B$'' of frequency $w_B$.
\begin{equation}
x_B(t) = \sqrt{\frac{2}{m_Bw_B}} \alpha_0 \cos w_Bt~.
\label{eq-amp}
\end{equation}
Here $\alpha_0$ is a real number representing the initial amplitude of this classical harmonic oscillator.
Note that we have set the mass of oscillator $A$ to 1, but we are keeping the mass of oscillator $B$ as a parameter.
It will be a very convenient variable to adjust for us.
Putting this back into Eq.~(\ref{eq-HA}), we get
\begin{eqnarray}
H_A &=& \frac{P_A^2}{2} + 
\left(\frac{w_0^2+w_A^2}{4} + \frac{w_0^2-w_A^2}{4}
\frac{x_B(t)}{\sqrt{\frac{2}{m_B w_B}}\alpha_0}\right)X_A^2~.
\label{eq-HA1}
\end{eqnarray}
Later in this paper, whenever we adjust $\alpha_0$, we will adjust $m_B$ together such that $\alpha_0^2/m_B = const.$
With this condition enforced, $\alpha_0$ is basically a free parameter for the classical model.
Both the classical amplitude of oscillator $B$ and the coupling between the two systems are unchanged when we vary $\alpha_0$.
Any value of $\alpha_0$ leads to the same time-dependent Hamiltonian for oscillator $A$.

Now, in order to study the unitarity problem, we should treat both oscillators quantum-mechanically.
That means they actually together follow a time-independent Hamiltonian.
\begin{equation}
H_{tot} = \frac{P_A^2}{2}+ \frac{P_B^2}{2m_B} + \frac{m_B w_B^2 X_B^2}{2} + 
\left( \frac{w_0^2+w_A^2}{4}+ \frac{w_0^2-w_A^2}{4} \frac{X_B}{\sqrt{\frac{2}{m_B w_B}}\alpha_0}\right)X_A^2~.
\label{eq-HAB}
\end{equation}
The classical amplitude, Eq.~(\ref{eq-amp}), is replaced by the position operator acting on a coherent state of the oscillator $B$.
\begin{equation}
|\phi_B(0)\rangle = e^{-\alpha_0^2/2} 
\sum_n \frac{\alpha_0^n}{\sqrt{n!}}|n\rangle_B~.
\label{eq-Bcoh}
\end{equation}
If the energy in oscillator $B$ is much larger than the combined energy in oscillator $A$ and the coupling term, then $\langle X_B \rangle$ will evolve like Eq.~(\ref{eq-amp}).

The combination of Eq.~(\ref{eq-HAB}) and (\ref{eq-Bcoh}) represents a 1-parameter family of fully quantum models parametrized by $\alpha_0$.
They all correspond to the same semi-classical model in Eq.~(\ref{eq-HA}).
However, the quantum realization of oscillator $B$ depends on $\alpha_0$.
One key difference between quantum and classical oscillators is that classical amplitudes are exact, but quantum expectation values have uncertainties.
The parameter $\alpha_0$ is the unit-less ratio between the amplitude and its uncertainty.
\begin{equation}
\alpha_0 \sim \frac{\langle X_B \rangle}{\langle \Delta X_B\rangle}~.
\label{eq-classical}
\end{equation}
Based on this property, it is probably easy to guess what values of $\alpha_0$ make the semi-classical approximation better.
We will soon see that indeed, the larger $\alpha_0$ is, the semi-classical model stays longer as a good approximation to the fully quantum model.

The unitarity question in the semi-classical model is simply an evolution problem in this fully quantum model.
If we start with a product state:
\begin{equation}
|\psi(0)\rangle_{AB} = 
\left( \sum_m c_m |m\rangle_A \right)
\left( e^{-\alpha_0^2/2} 
\sum_n \frac{\alpha_0^n}{\sqrt{n!}}|n\rangle_B \right)~,
\label{eq-product}
\end{equation}
how long will they stay as an approximate product state? 
When will they become significantly entangled with each other?

\subsection{Non-degenerate Perturbation Theory}
\label{sec-ndpt}

One simple way to solve the evolution of the combined system is to treat it as a time-independent perturbation theory.
The unperturbed Hamiltonian is for two uncoupled harmonic oscillators.
Oscillator $B$ has frequency $w_B$, while oscillator $A$ has the mean frequency $\bar{w} = \sqrt{(w_0^2+w_A^2)/2}$.
\begin{eqnarray}
H_0 = \bar{w}
\left(a^\dagger a+\frac{1}{2}\right) + w_B\left(b^\dagger b + \frac{1}{2}\right)~.
\end{eqnarray}
Here $a$ and $b$ are the standard lowering operator for oscillators $A$ and $B$ respectively.
\begin{eqnarray}
a &=& \frac{1}{\sqrt{2}} \left(\bar{w}^{1/2}X_A+i \bar{w}^{-1/2}P_A \right)~, \\
b &=& \frac{1}{\sqrt{2}} \left[ (m_Bw_B)^{-1/2} X_B + i (m_B w_B)^{-1/2} P_B \right]~.
\end{eqnarray}
The eigenstates for the unperturbed Hamiltonian are product states of the eigenstates of the individual oscillators.
\begin{equation}
H_0 |m\rangle_A |n\rangle_B 
= \left[ \bar{w}\left(m+\frac{1}{2}\right) + w_B\left(n+\frac{1}{2}\right)\right]
|m\rangle_A |n\rangle_B
\equiv E^{(0)}_{mn} |m\rangle_A |n\rangle_B~.
\end{equation}

The perturbation is the coupling between them.
\begin{eqnarray}
V = \frac{w_0^2-w_A^2}{4}\sqrt{\frac{w_B}{2}}\frac{X_B}{\alpha_0}X_A^2
= \epsilon\left(a+a^\dagger\right)^2\left(b+b^\dagger\right)~.
\end{eqnarray}
Here $\epsilon = \frac{w_0^2-w_A^2}{16\bar{w}\alpha_0}$ will be a small number when $\alpha_0$ is large.
In addition, if we $\bar{w}$ and $w_B$ are incommensurate, there will be no degeneracy in the unperturbed spectrum.
It appears that we can solve it as a standard non-degenerate, time-independent perturbation theory.

The energy eigenstates, corrected up to the first order, becomes
\begin{equation}
|\psi_{mn}\rangle = |m\rangle_A |n\rangle_B + 
\epsilon \sum_{p,q} \frac{\langle p |\left(a+a^\dagger\right)^2|m\rangle_A
\langle q| \left(b+b^\dagger\right) |n\rangle_B }
{\bar{w}(m-p)+w_B(n-q)} |p\rangle_A|q\rangle_B~.
\label{eq-EigenStateCor}
\end{equation}
The eigenstate energy is corrected at the second order.
\begin{eqnarray}
\Delta E_{mn} =
\label{eq-EnergyCor}
\epsilon^2 \sum_{p,q} \frac{\bigg|\langle p |\left(a+a^\dagger\right)^2|m\rangle_A
\langle q| \left(b+b^\dagger\right) |n\rangle_B\bigg|^2 }
{\bar{w}(m-p)+w_B(n-q)}~.
\end{eqnarray}
The summation of $p,q$, in both cases, only run through 6 states: $p=(m+2), m, (m-2)$ and $q=(n+1), (n-1)$.
Thus it is easy to calculate
\begin{eqnarray}
\epsilon^{-2} \Delta E_{mn} &=& \frac{(m+1)(m+2)(n+1)}{-2\bar{w} - w_B}
+ \frac{(2m+1)^2(n+1)}{-w_B} 
\label{eq-DeltaE}
\\ \nonumber
&+& \frac{m(m-1)(n+1)}{2\bar{w}-w_B}
+ \frac{(m+1)(m+2)n}{-2\bar{w} + w_B}
+ \frac{(2m+1)^2n}{w_B} + \frac{m(m-1)n}{2\bar{w}+w_B} 
\\ \nonumber
&=& \frac{m^2-4mn-m+2n}{2\bar{w}-w_B} - \frac{m^2+4mn+3m+2n+2}{2\bar{w}+w_B}
-\frac{(2m+1)^2}{w_B}~.
\end{eqnarray}

In order to describe the evolution, we can first rewrite Eq.~(\ref{eq-product}) in the perturbed eigenstate bases.
\begin{eqnarray}
|\psi(0)\rangle = \sum_{m,n} c_m  e^{-\alpha_0^2/2} \frac{\alpha_0^n}{\sqrt{n!}}
\left[1 + \mathcal{O}(\epsilon) \right]|\psi_{mn}\rangle
 \end{eqnarray}
Here the $\mathcal{O}(\epsilon)$ comes from the corrections to the eigenstates in Eq.~(\ref{eq-EigenStateCor}).
They are small real numbers suppressed by $\epsilon$.
The time evolution is then simply a phase in the eigenstate basis.
\begin{eqnarray}
|\psi(t)\rangle &=& \sum_{m,n} e^{-i E_{mn}t} c_m  e^{-\alpha_0^2/2} \frac{\alpha_0^n}{\sqrt{n!}}
\left[1 + \mathcal{O}(\epsilon) \right]|\psi_{mn}\rangle 
\\ \nonumber 
&=& \sum_{m,n} e^{-i E_{mn}t} c_m  e^{-\alpha_0^2/2} \frac{\alpha_0^n}{\sqrt{n!}}
\left[1 + \mathcal{O}(\epsilon) \right]|m\rangle_A|n\rangle_B~.
\end{eqnarray}
We used Eq.~(\ref{eq-EigenStateCor}) again to put it back into product state basis.
The values of $\mathcal{O}(\epsilon)$ has changed.
They are still small, but they are complex now due to the different phases.
This small correction is actually not important, because we only want to check whether it is still a product state.

For example, let us assume that initially, only $m=0$ and $2$ states are excited for oscillator $A$; $c_0=c_2=1/\sqrt{2}$ and all other $c_m=0$.
We can check whether the two oscillators are entangled by first partially projecting oscillator $A$ into these two states. 
\begin{eqnarray}
|A_0\rangle_B \equiv \langle 0|_A|\psi(t)\rangle \bigg|_{\rm normalized} 
&=& \sum_n e^{-i\Delta E_{0n}t}\frac{\alpha_0^n}{\sqrt{n!}}
\left[1 + \mathcal{O}(\epsilon) \right] e^{-iE^B_nt} |n\rangle_B~, 
\label{eq-repeatStart} \\
|A_2\rangle_B \equiv \langle 2|_A|\psi(t)\rangle \bigg|_{\rm normalized} 
&=& \sum_n e^{-i\Delta E_{2n}t}\frac{\alpha_0^n}{\sqrt{n!}}
\left[1 + \mathcal{O}(\epsilon) \right] e^{-iE^B_nt} |n\rangle_B~.
\end{eqnarray}
If these two corresponding states in $B$ are orthogonal, then a projection in $A$ also works as a projection in $B$.
That shows the two oscillators are significantly entangled
\footnote{Particle production effect will start to excite higher states in oscillator $A$. 
Thus, just checking the $m=0$ and $m=2$ projections cannot show that they are maximally entangled.
Nevertheless, if the frequencies are far away from being resonant, higher states will not be excited by a lot.
These two lowest states is sufficient to show that they are significantly entangled.}.
On the other hand, if these two corresponding states in $B$ are parallel, then a projection in $A$ has no effect on $B$.
That means we still have a product state, thus the unitarity on $A$ alone is still valid.

The answer then depends on the inner product between these two states, which is a sum of complex numbers.
\begin{eqnarray}
\langle A_0|A_2\rangle 
 &\propto& \sum_n e^{i(\Delta E_{2n}-\Delta E_{0n})t}
 \frac{\alpha_0^{2n}}{n!}
\left[1 + \mathcal{O}(\epsilon) \right]~.
\label{eq-repeatEnd}
\end{eqnarray}
This sum of $n$ is dominated by a range between $n_{min}\sim(\alpha_0^2-\alpha_0)$ and $n_{max} \sim (\alpha_0^2+\alpha_0)$.
So the basic question here is:
{\it Within this range, do we have small phase differences among the terms in the above sum?} 
If the differences are small, then all terms add up coherently, and the inner product is close to one, which suggests that the two states are parallel. 
On the other hand, if the differences are large, then they will cancel each other in the sum, which suggest that the two states are orthogonal.

From Eq.~(\ref{eq-EnergyCor}), we can estimate this phase change.
\begin{eqnarray}
& & (\Delta E_{2n_{max}}- \Delta E_{0n_{max}})t - 
(\Delta E_{2n_{min}}- \Delta E_{0n_{min}})t
\label{eq-phase}
\\ \nonumber 
 &\approx& \epsilon^2
\left[ \frac{-8(n_{max}-n_{min})}{2\bar{w}-w_B} -\frac{8(n_{max}-n_{min})}{2\bar{w}+w_B} \right]t
\\ \nonumber
&=&-16\alpha_0\epsilon^2 \frac{4\bar{w}}{4\bar{w}^2-w_B^2}t
= -\frac{1}{4\alpha_0}\frac{(w_0^2-w_A^2)^2}{\bar{w}(4\bar{w}^2-w_B^2)}t~.
\end{eqnarray}

This reveals a critical time scale after which the two oscillators become entangled.
\begin{equation}
T_{ent} \sim 8\pi \alpha_0 
\frac{ \bar{w} (4\bar{w}^2-w_B^2) }{(w_0^2-w_A^2)^2}~.
\label{eq-entangle}
\end{equation}
This is the main technical result of this paper.
At this time scale, the initial pure state of oscillator $A$ becomes a mixed state.
The unitarity for this oscillator alone is lost.

Recall that a coherent state has the same uncertainty as the ground state, 
\begin{equation}
\langle \alpha_0 | \Delta X_B |\alpha_0\rangle = \sqrt{\frac{2}{w_B}}~.
\end{equation}
We see that $\alpha_0 = \langle X_B\rangle/\langle \Delta X_B\rangle$ is a good parameter to quantify how ``classical'' the oscillator $B$ is.
This of course makes sense.
The more ``classical'' the background is, the longer can oscillator $A$ stays unitary on its own.
However, our point here is that from the semi-classical point of view, the parameter $\alpha_0$ is an extra free parameter.
All values of $\alpha_0$ basically leads to the same semi-classical model.
Thus, there is no way, from the semi-classical model alone, to foresee this loss of unitarity.

\subsection{The need of numerical confirmation}

Although Eq.~(\ref{eq-entangle}) seems to be a very reasonable result, the derivation in the previous section cannot be used as a rigorous proof.
Non-degenerate perturbation theory requires the condition that $\Delta E_{mn}$ is small compared to the unperturbed gaps of eigen-energies. 
We can see that such requirement is not satisfied if we take a closer look at Eq.~(\ref{eq-DeltaE}).
For any pair of fixed $(m,n)$, $\Delta E_{mn}$ does decrease with $\epsilon^2\propto\alpha_0^{-2}$.
However, we should recall that the relevant values of $n$ also change with $\alpha_0$.
In fact, the expectation value of energy level of a coherent state is proportional to $\alpha_0^2$.
Such dependence exactly cancels the $\epsilon$ suppression.
Thus for all relevant eigenstates $(m,n)$ in this calculation, $\Delta E_{mn}$ is comparable to the gaps of $E^{(0)}_{mn}$.
Consequently, the perturbative expansion of eigenstates in Eq.~(\ref{eq-EigenStateCor}) is also questionable.
For a smaller value of $n$, the non-degenerate perturbation theory is justified.
But as we increase to $n\sim\alpha_0^2$, it starts to break down.

In order to get the exact answer, we need degenerate perturbation theory.
Namely, we have to diagonalize the Hamiltonian without assuming it as a small perturbation from the unperturbed one.
We will do that numerically in the next section.
Here we will first point out that the results agree with Eq.~(\ref{eq-entangle}), and there is a very good reason.

If we trace back the derivation to Eq.~(\ref{eq-phase}), we can see that our conclusion did not care about the absolute correction to the energy $\Delta E_{mn}$.
It cares about the difference between two of such corrections, namely $(\Delta E_{mn_1}-\Delta E_{mn_2})$, with $|n_1-n_2|\sim\alpha_0$.
This value is small comparing to the energy gaps.
Since it is the fundamental reason behind our physical conclusion, it is not too surprising that the answer is correct.
\footnote{
Although, it is very tempting to look for the exact analytical proof for that.
We will leave that to future work.}

\section{Numerical Method}
\label{sec-numeric}

In this section, we will show two examples in which all the physical parameters are chosen to be some numerical values.
We can then directly diagonalize the Hamiltonian in Eq.~(\ref{eq-HAB}) by Mathematica.
Of course, the actual Hilbert space is infinite-dimensional, so we will truncate it down to a finite size.
First of all, we will start with $m=0$ and $m=2$, and the frequencies are not close to resonance.
That means we do not expect higher eigenstates of oscillator $A$ to be populated significantly.
We will have a cutoff $m_{Max}$ at 6 or 8 and make sure that indeed the highest state has little influence on the result.
For oscillator $B$, we limit ourselves between $n_{Min} = (\alpha_0^2-\kappa\alpha_0)$ and $n_{Max}= (\alpha_0^2+\kappa\alpha_0)$.
For the results presented in this paper, we use $\kappa=8$ which we checked that a higher value no longer changes the outcome.

In this truncated Hilbert space, we diagonalize the Hamiltonian and replace Eq.~(\ref{eq-EigenStateCor}) by the actual transformation to the true eigenstates.
\begin{eqnarray}
|\psi_{pq}\rangle &=& \sum_{m,n} \Lambda_{pq}^{mn}|m\rangle_A|n\rangle_B ~, \\
H_{tot}|\psi_{mn}\rangle &=& E_{mn}|\psi_{mn}\rangle~.
\end{eqnarray}
Note that $|\psi_{mn}\rangle$ is no longer close to $|m\rangle|n\rangle$, but the total number of eigenstates does not change.
Namely, the first subscript of $\psi$ still runs through the range of $m$, and the second subscript still runs through the range of $n$.

The time evolution is also modified to
\begin{eqnarray}
|\psi(t)\rangle &=& \sum_{m,n} c_m e^{-\alpha_0^2/2} \frac{\alpha_0^n}{\sqrt{n!}} 
\sum_{pq} (\Lambda^{-1})_{mn}^{pq} e^{-iE_{pq}t} |\psi_{pq}\rangle
\\ \nonumber
&=&\sum_{m,n} c_m e^{-\alpha_0^2/2} \frac{\alpha_0^n}{\sqrt{n!}} 
\sum_{pq} (\Lambda^{-1})_{mn}^{pq} e^{-iE_{pq}t} 
\sum_{r,s} \Lambda_{pq}^{rs} |r\rangle_A|s\rangle_B
\\ \nonumber
&=& \sum_r |r\rangle_A \sum_s
\bigg( \sum_{m,n}\sum_{p,q}c_m e^{-\alpha_0^2/2} \frac{\alpha_0^n}{\sqrt{n!}} 
(\Lambda^{-1})_{mn}^{pq} e^{-iE_{pq}t}  \Lambda_{pq}^{rs}\bigg)
|s\rangle_B~.
\end{eqnarray}
We then use this to repeat the calculation from Eq.~(\ref{eq-repeatStart}) to (\ref{eq-repeatEnd}).

\subsection{The Classical Limit}
\label{sec-CL}

In our first example, we will use a fixed semi-classical model with the following parameters:
$w_0=2$, $w_A=1$, $w_B=0.5$.
The only parameter we will change is $\alpha_0$. 
Different values of $\alpha_0$ correspond to different quantum models which all have the same semi-classical approximation.
Eq.~(\ref{eq-entangle}) tells us that the larger $\alpha_0$ is, the longer can oscillator $A$ stay unitary on its own.

\begin{figure}[tb]
\begin{center}
\includegraphics[scale = 0.8]{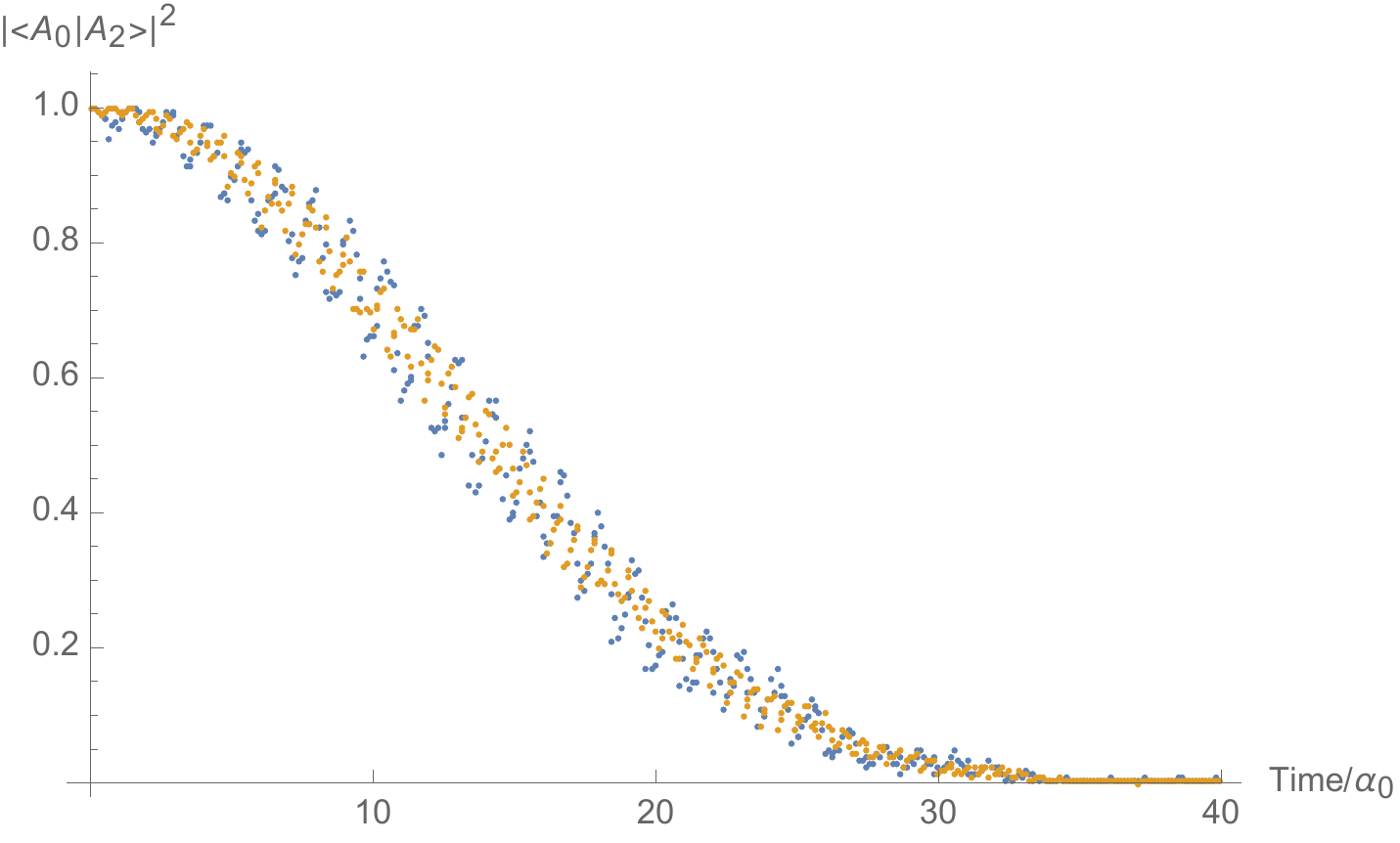}
\caption{The value of $|\langle A_0|A_2\rangle|^2$ as a function of time, rescaled by $\alpha_0$.
The blue dots are for $\alpha_0=10$, and the orange dots for $\alpha_0=20$.
The other parameters used here are $w_0=2$, $w_A=1$, $w_B=(1/2)$.
}
\label{fig-CL}
\end{center}
\end{figure}

We numerically solve $\langle A_0|A_2\rangle$ as a function of time and plot it in Fig.\ref{fig-CL}.
Plugging the values of $w_0$, $w_A$ and $w_B$ to Eq.~(\ref{eq-entangle}), we expect that after
\begin{equation}
T_{ent} \sim 43\alpha_0~,
\end{equation}
this inner-product should approach zero, 
and oscillator $A$ will be significantly entangled with oscillator $B$.
We present the results of two different values of $\alpha_0$, and it shows that such prediction is quite accurate.
The more ``classical'' oscillator $B$ is, the longer will oscillator $A$ remain unitary on its own.

\begin{figure}[tb]
\begin{center}
\includegraphics[scale = 0.85]{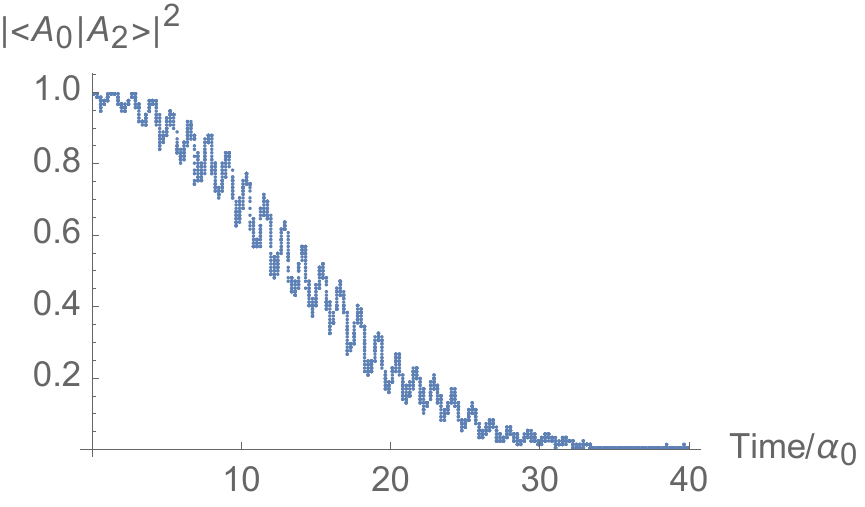}
\includegraphics[scale = 0.85]{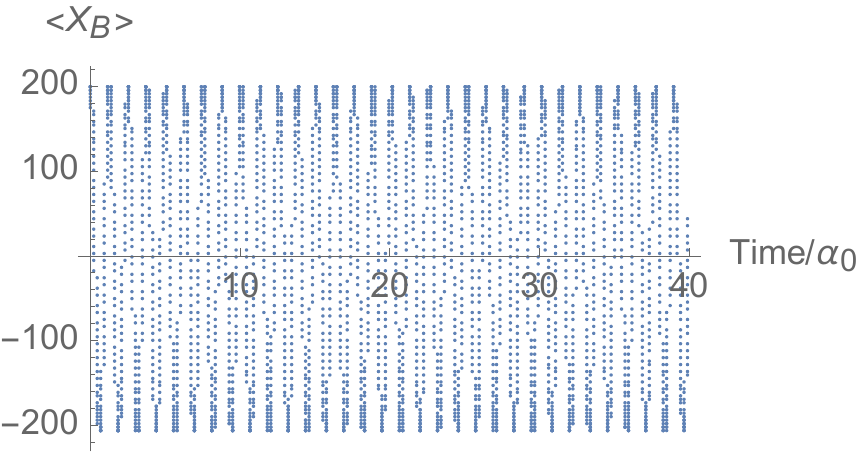}
\caption{
Left: The same function $|\langle A_0|A_2\rangle|^2$ for $\alpha_0=10$ in higher time-resolution.
Right: The value of $\langle X_B\rangle$ as a function of time.
}
\label{fig-highRes}
\end{center}
\end{figure}

The readers might notice that the spread of the curves in Fig.\ref{fig-CL} is different between two different values of $\alpha_0$, and seems to have a pattern.
Indeed, in Fig.\ref{fig-highRes} we plot in higher resolution and see that $|\langle A_0|A_2\rangle|^2$ does not follow a monotonic curve.
It has small oscillations at the same frequency as the amplitude of oscillator $B$.
Our next example will shed more light on that.
Here we will can see from Fig.\ref{fig-highRes} that the classical motion of oscillator $B$ is barely affected while oscillator $A$ loses its unitarity.
Showing that such loss of unitarity is an independent effect from classical back-reactions.

\begin{figure}[tb]
\begin{center}
\includegraphics[scale = 0.85]{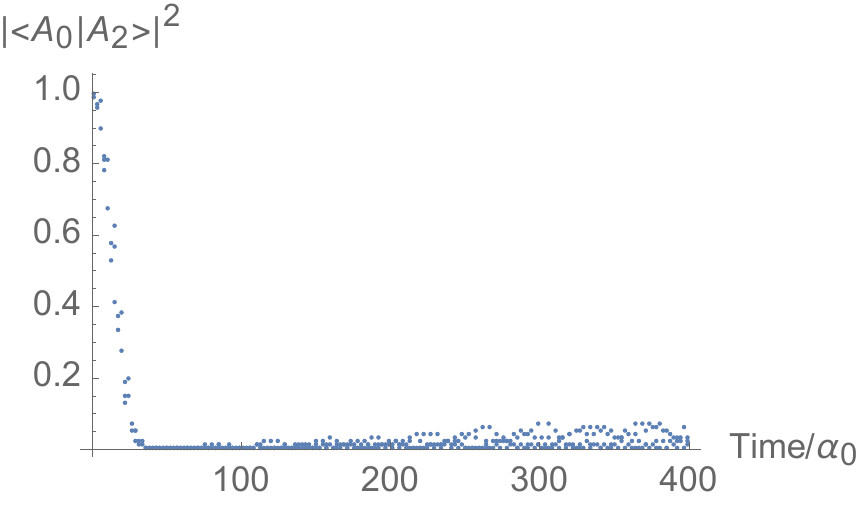}
\includegraphics[scale = 0.85]{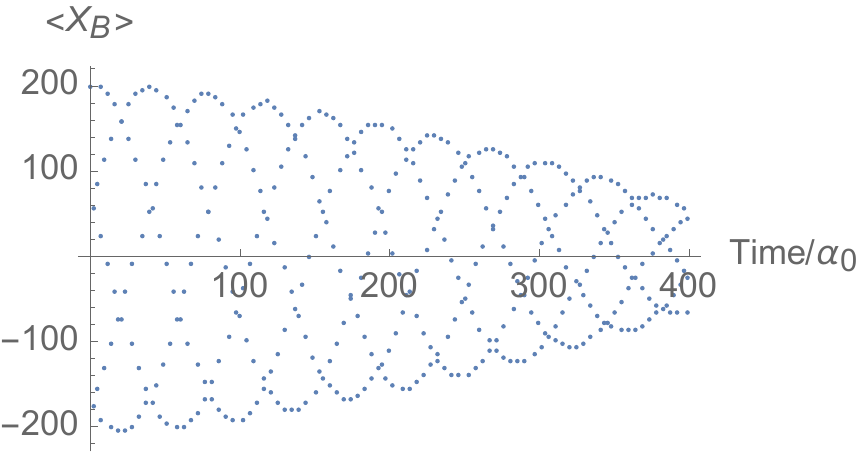}
\caption{
Left: The same function $|\langle A_0|A_2\rangle|^2$ for $\alpha_0=10$ for longer time.
Right: The value of $\langle X_B\rangle$ for longer time.
}
\label{fig-long}
\end{center}
\end{figure}

One may wonder whether such loss of unitarity is permanent.
After all, the decreasing behaviour of $|\langle A_0|A_2\rangle|^2$ up to $t\sim T_{ent}$ could have been part of a sinusoidal function, which could come back to $1$ at $t\sim 2T_{ent}$.
In Fig.\ref{fig-long} we extend Fig.\ref{fig-highRes} to longer time to show that the value of $|\langle A_0|A_2\rangle|^2$ stays near zero thereafter.
It is worth noting that on this longer time scale, back-reaction to the motion of oscillator $B$ becomes significant.
That undermines the validity of our truncated Hilbert space in the numerical approach.
Nevertheless, both back-reaction and spreading out more in the Hilbert space are physical reasons to support even further loss of unitarity, instead of any miraculous restoration of unitarity.

Based on the analytical result, we expect the two oscillators to stay entangled as long as the complex terms in Eq.~(\ref{eq-repeatEnd}) have incoherent phases.
Therefore, they will only become unentangled when all the phases grow to multiples to $2\pi$.
That happens when $t\sim \alpha_0 T_{ent}$, which is basically the recurrence time of this system.
Thus we are confident to say that the unitarity of oscillator $A$ alone is irrevocably lost until the much longer time scale.

\subsection{The Adiabatic Limit}
\label{sec-AL}

In the previous section we have shown that a gedanken ``classical limit'' exists.
One can imagine a family of quantum models that correspond to the same semi-classical approximation, and they make the approximation better and better as we tune some variable.
That however, is not always the relevant answer to important question in practice.
In the case of QFT-CST, we usually assume that the semi-classical approximation is about one unique quantum theory of gravity.
The interesting physical question is applying such approximation to different semi-classical situations.

For example, while formulating the black hole information paradox, it is customary to take the ``large black hole'' limit.
In this limit, the geometry (classical background) changes much slower than the natural frequency of the QFT modes we are analyzing (Hawking radiation).
That is basically the adiabatic limit, in which the semi-classical background varies slowly, $w_B\ll \bar{w}$.
We will explore the behaviour in such limit in this section by reducing the value of $w_B$ further.

Interestingly, we cannot simply reduce $w_B$ while holding $\alpha_0$ fixed.
The total energy of oscillator $B$ is about $(w_B\alpha_0^2)$, which gets smaller with $w_B$ if $\alpha_0$ is fixed.
As the total energy in oscillator $B$ gets smaller, the back-reactions become more significant, which is not the situation we would like to analyze.
We would like to ensure that the classical background remains ignorant to back-reactions.
Therefore, we will hold the energy of oscillator $B$ constant as we reduce $w_B$.
That means $\alpha_0\propto w_B^{-1/2}$.

From the analytical results, we can see that Eq.~(\ref{eq-entangle}) stops depending explicitly on $w_B$ when it gets small.
There is still an implicit dependence through $\alpha_0$.
\begin{equation}
T_{ent} \propto \alpha_0 \propto w_B^{-1/2}~.
\end{equation}
This does get longer as $w_B$ decreases, but the relevant physical question is to compare it to the time scale in which the background changes in the semi-classical model.
For example, in order for the effective frequency of oscillator $A$ to go from $w_0$ to $w_A$, it takes time $\pi w_B^{-1}$.
In other words, the natural time scale of the change in classical background is $T_{background}\sim w_B^{-1}$.

\begin{figure}[tb]
\begin{center}
\includegraphics[scale = 1]{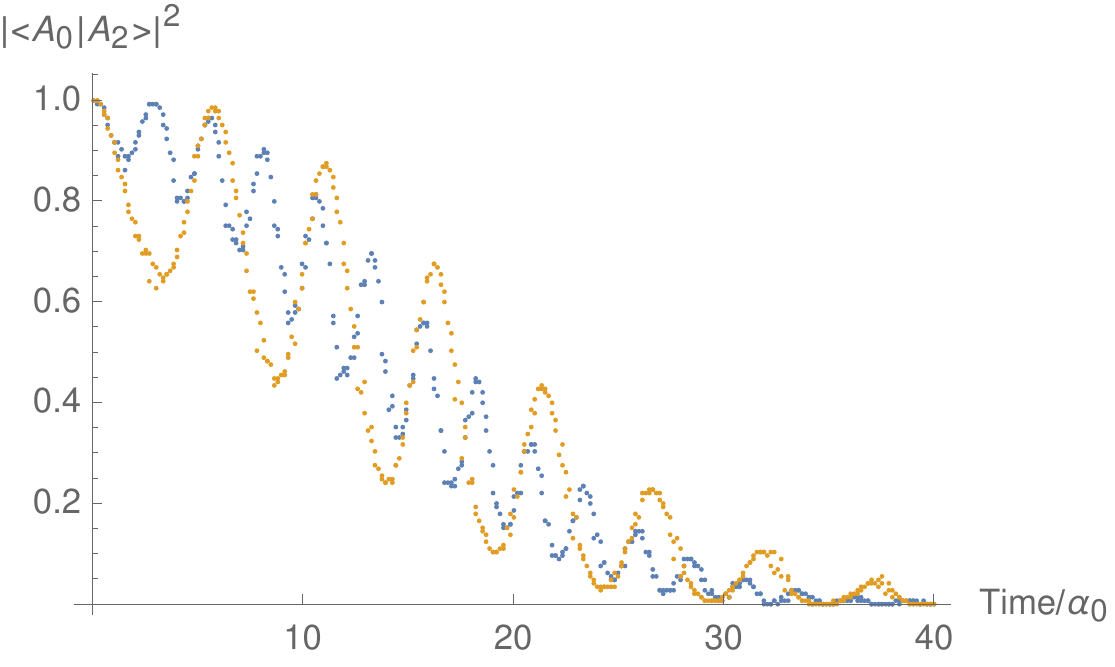}
\caption{The value of $|\langle A_0|A_2\rangle|^2$ as a function of time, rescaled by $\alpha_0$.
The blue dots are for $\alpha_0=20$ and $w_B=(1/4)$.
The orange dots are for $\alpha_0=40$ and $w_B=(1/16)$.
The other parameters used here are $w_0=2$, $w_A=1$.
}
\label{fig-adiabatic}
\end{center}
\end{figure}

When $w_B\rightarrow0$, we are guaranteed to have $T_{ent}\ll T_{background}$. 
Thus {\bf the unitarity of oscillator $A$ alone is always lost in this adiabatic limit!}
\footnote{Note that this is related to the fact that we fixed the total energy in oscillator $B$ while taking the limit.
So it is an example that the loss of unitarity can happen, not a proof that it generically will happen.}
Fig.\ref{fig-adiabatic} shows the numerical result as we reduce $w_B$ to support this conclusion.

\begin{figure}[tb]
\begin{center}
\includegraphics[scale = 0.85]{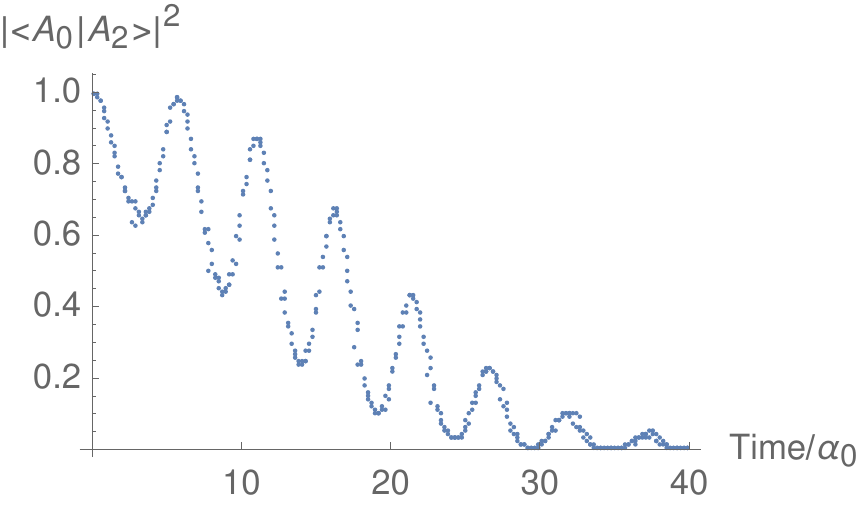}
\includegraphics[scale = 0.85]{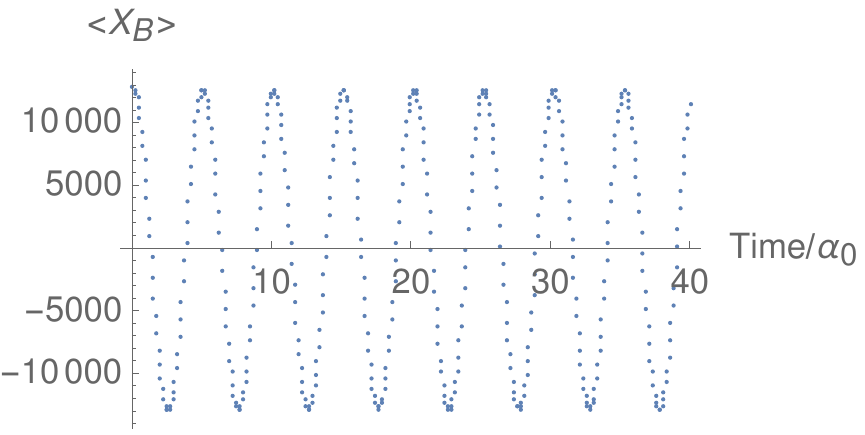}
\caption{
Left: The same function $|\langle A_0|A_2\rangle|^2$ for $\alpha_0=40$ and $w_B=(1/16)$.
Right: The value of $\langle X_B\rangle$.
}
\label{fig-oscillate}
\end{center}
\end{figure}

We can see that the overall decrease of $|\langle A_0|A_2\rangle|^2$ is the same as the irrevocable loss of unitarity as we shown in the previous section.
On top of that, the oscillating behaviour becomes more obvious and prominent.
We again plot it together with the amplitude of oscillator $B$ in Fig.\ref{fig-oscillate}.
Just like in our previous examples, the back-reaction has no visible effect on the classical motion of oscillator $B$, but $|\langle A_0|A_2\rangle|^2$ is clearly being modulated by $\langle X_B\rangle$.
We will provide the analytics explanation of such behaviour in the next section.

\section{Revocable and Irrevocable Losses of Unitarity}
\label{sec-Losses}

In Fig.\ref{fig-highRes} and \ref{fig-oscillate}, we can clearly see that in addition to the long-term, irrevocable loss of unitarity, there is a short-term oscillation.
Such oscillation becomes more prominent in the adiabatic limit where $w_B\ll \bar{w}$.
When $\langle X_B \rangle$ evolves to furthest away from its initial value, the two oscillators get very entangled.
When $\langle X_B \rangle$ evolves back to the initial value, the two oscillators get less entangled.
It suggests that there is a different kind of unitarity loss that is periodic and revocable to a certain extent.
Here we will provide the analytical explanation of such phenomenon.
We will show that it is different and happens on top of the irrevocable loss.

Such phenomenon is easiest to address in a slightly different basis.
Previously, for oscillator $A$, we have been using a time-independent basis $|m\rangle$ which is the occupation-number basis for its mean frequency $\bar{w}$.
In the adiabatic limit, a time-dependent basis, $|m\rangle_{w(t)}$, according to the occupation number of its instantaneous frequency $w(t)$, is far more convenient.
That is because in the semi-classical model, no particles will be produced.
That means the occupation number for the instantaneous frequency stays the same.

Strictly speaking, we will analyze a slightly different problem.
The particular initial state in the $\bar{w}$ basis,
\begin{equation}
|\phi_A(0)\rangle = \frac{1}{\sqrt{2}} \left( |0\rangle_{\bar{w}} + |2\rangle_{\bar{w}} \right)~,
\end{equation}
will have nonzero components in more than just $|0\rangle_{w(t)}$ and $|2\rangle_{w(t)}$.
Nevertheless, as long as $w_0$ and $w_A$ are not too different, it will still be dominated by low and even eigenstates.
So we can still monitor the entanglement by focusing on these two eigenstates.
In other words, we might have just chosen a different initial condition.
\begin{equation}
|\phi_A(0)\rangle = \frac{1}{\sqrt{2}} \left( |0\rangle_{w(0)} + |2\rangle_{w(0)} \right)~.
\end{equation}
This is similar enough to the previous initial condition, and such small change do not affect the main conclusion.

Other than the slightly different initial condition, there is nothing else we need to worry about.
Entanglement is an intrinsic property between the two oscillators, which will not change if we analyze the problem in a different basis.

The time evolution of the combined system, in this basis, can be written simply as
\begin{eqnarray}
|\psi(t)\rangle_{AB} &=& \frac{1}{\sqrt{2}} 
\left( |0\rangle_{w(t)}|A_0(t)\rangle_B +|2\rangle_{w(t)}|A_2(t)\rangle_B   \right)~.
\end{eqnarray}
$|A_0\rangle$ and $|A_2\rangle$ started as the same coherent state.
\begin{equation}
|A_0(0)\rangle = |A_2(0)\rangle = |\alpha_0\rangle \equiv
e^{-\alpha_0^2/2} \sum_{n\sim(\alpha_0^2-\alpha)}^{n\sim (\alpha_0^2+\alpha)} 
\frac{\alpha_0^n}{\sqrt{n!}}|n\rangle_B~.
\label{eq-range}
\end{equation}
The question is how $\langle A_0|A_2\rangle$ evolves with time afterward.

Note that we deliberately show the effective range on the summation of $n$.
A coherent state $\alpha_0$ has the energy expectation value $E \sim w_B\alpha_0^2$ because the effective range enters at $n\sim\alpha_0^2$.
It has an energy uncertain $\Delta E \sim w_B\alpha_0$ because it is dominantly the superposition of energy eigenstates within a range $\Delta n\sim\alpha_0$.

Both the total energy and the total energy uncertainty are conserved quantities.
The state of $A$, $|n\rangle_{w(t)}$, has a time-dependent energy but no energy uncertainty.
That means without knowing other details of $|A_n(t)\rangle_B$, we know that it must retain the same energy uncertainty from the beginning, but has an opposite time-dependence in the mean energy to compensate the change in $A$.
In other words, when $w(t)$ evolves from $w_0$ to $w_A$, the corresponding state in $B$ has to be made from eigenstates in a shifted range of the same size.
\begin{equation}
|A_m(\pi w_B^{-1})\rangle_B = \sum_{n\sim[\alpha_0^2-\alpha_0 + mw_B^{-1}(w_0-w_A)]}^
{n\sim[\alpha_0^2 + \alpha_0 + mw_B^{-1}(w_0-w_A)]}
c_{mn} |n\rangle_B~.
\label{eq-rangeshift}
\end{equation}

Thus, if $w_B^{-1}(w_0-w_A) > \alpha_0$, we must have $\langle A_0(\pi w_B^{-1})|A_2(\pi w_B^{-1})\rangle\sim0$, since they do not overlap in this basis. 
This is exactly the effect of ``energy carries information'' as discussed in \cite{IlgYan14}.
We do not even need to consider the phases of $c_{mn}$ to see that the two states will become orthogonal.
Thus, in addition to calculating the phases as we did earlier, this is an independent reason why the two oscillators can become entangled.

When $t=2\pi w_B^{-1}$, $w(t)$ goes back to the initial value $w_0$, and whether these two states are orthogonal follows our calculation of phases in Sec.\ref{sec-ndpt}.
If the phases are still coherent, then the two oscillators again become unentangled.
The unitarity was only temporarily lost and is revocable
\footnote{This is very similar to the situation in the Stern-Gerlach experiment that we split and then merge the two jets of particles.}.
However, after a longer time, the effect that decoheres the phases kicks in, and the unitarity of individual oscillators will be irrevocably lost.

\begin{figure}[tb]
\begin{center}
\includegraphics[scale = 0.5]{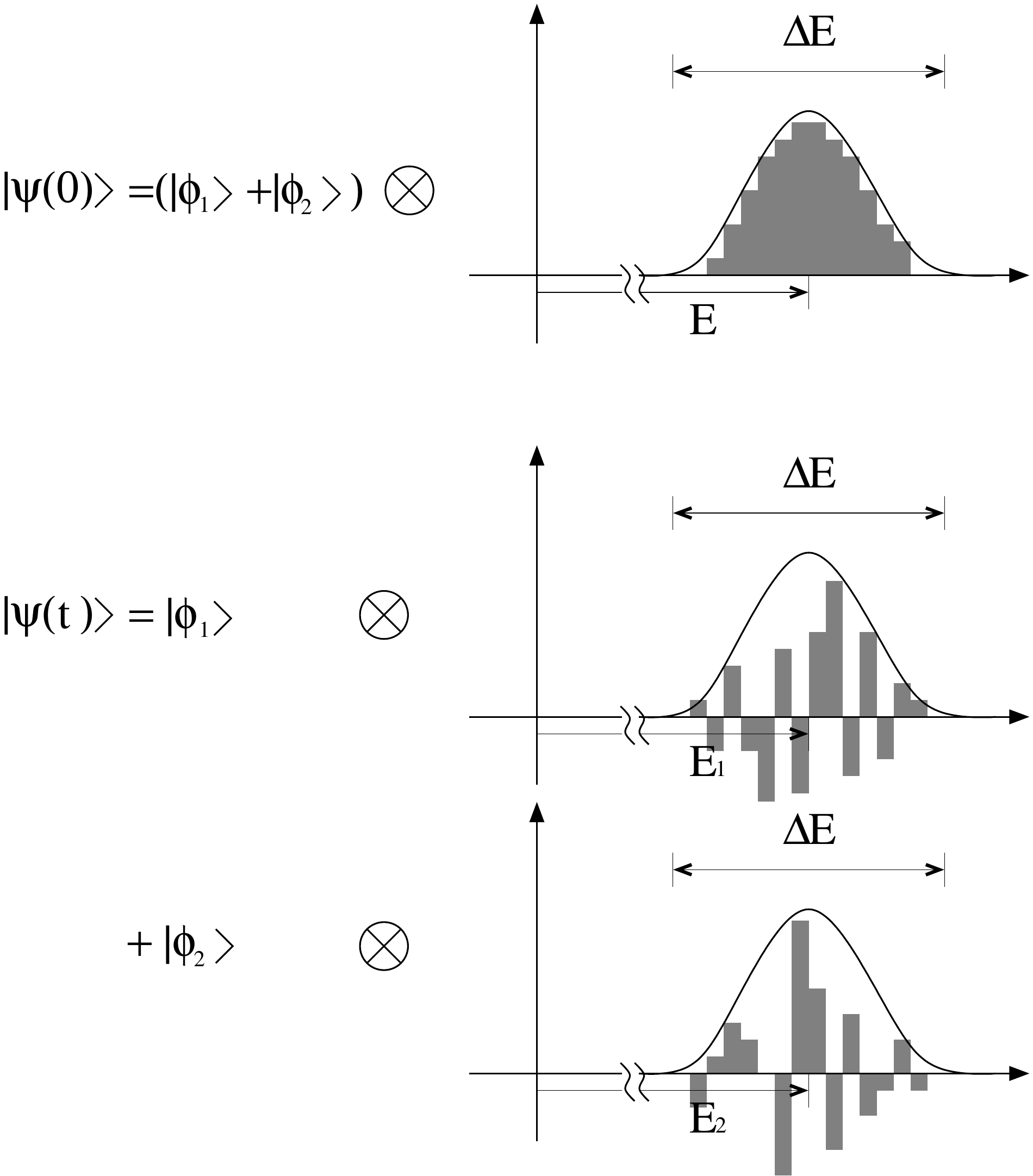}
\caption{
A cartoon for how the state of the classical background (oscillator $B$) evolves with time.
Its state is represented by the projections into energy-eigenstate-basis (grey bars).
It always stays as a very ``classical'' state, with energy uncertainty much less than the actual energy, $\Delta E \ll E$.
The energy shifts from back-reactions, $|E-E_1|$ and $|E-E_2|$, are both also much less than $E$.
However, if $|E_1-E_2|\gtrsim\Delta E$, clearly the two states become orthogonal, and the unitarity for the quantum subsystem (oscillator $A$) is secretly lost.
Even if $|E_1-E_2|\ll\Delta E$, the phases (heights and signs of the grey bars) can decohere, which will also make the two states orthogonal and destroy the subsystem unitarity.
}
\label{fig-states}
\end{center}
\end{figure}

In Fig.\ref{fig-states}, we drawn a cartoon to visualize the difference between revocable and irrevocable losses.
We represent the state of oscillator $B$ in Eq.~(\ref{eq-range}) and (\ref{eq-rangeshift}) as pictures of projections into the energy-eigenstate-basis.
We can see that these two types of losses are indeed independent and different effects.
It is easy to imagine that the same picture is not limited to our example.
Instead of the energy-eigenstate-basis, one can choose any other basis and apply this picture to all semi-classical approximations. 

We can see that in our first example in Sec.\ref{sec-CL}, as we increase $\alpha_0$, the uncertainty in energy also increases, thus the energy shift cannot lead to the revocable loss of unitarity.
In our second example in Sec.\ref{sec-AL}, we fixed the total energy in oscillator $B$, $E_B\sim w_B \alpha_0^2$.
That reduces the energy uncertainty, $\Delta E_B \sim w_B\alpha_0 $, which makes the revocable loss of unitarity more prominent.
At the same time, decreasing $w_B$ also delays the revocable loss.
Toward the adiabatic limit, we will have $w_B^{-1}>T_{end}$ sooner or later.
The irrevocable loss will not only occur before the revocable loss, it will occur even before any finite change in the effective frequency.
Therefore, the irrevocable loss can happen without any semi-classical back-reaction.

\section{Summary and Discussion}
\label{sec-dis}

In this paper, we provided simple examples to demonstrate that unitarity in semi-classical approximations can be lost without strong back-reactions.
We derived the critical time scale at which this secret loss of unitarity happens.
We numerically demonstrated two types of unitarity loss, and provided clear physical explanations for both.
The first type is revocable, associated with energy change, and it was discussed earlier in \cite{IlgYan14}.
The second type is irrevocable until the quantum recurrence time, and it is associated with the phase coherence of the semi-classical background.

Our result suggests that extra caution is required while invoking unitarity in semi-classical approximations.
Ideally, every semi-classical approximation should come with a consistency check to see whether its apparent unitarity is valid.
Our example shows that such consistency check may have to go beyond the semi-classical approximation itself, since we needed to explicitly use the underlying quantum theory.

The revocable loss is somewhat within the scope of the semi-classical approximation, since we are simply comparing energy shifts to energy uncertainty.
The energy shifts in Eq.~(\ref{eq-rangeshift}) and Fig.\ref{fig-states} can be calculated by back-reaction.
The energy uncertainty is intrinsically quantum, but we may also picture it as a classical quantity.
At the very least, we can imagine that it has a fixed (although unknown) value.
Thus if the energy shifts from the back-reaction goes to zero, we are guaranteed to have no unitarity loss of the revocable type.
This is the trick Feynman used his Lectures to explain the double-slit interference experiment
\footnote{See \cite{BakKod17} for a quick review.}.

However, the irrevocable type of unitarity loss is about quantum phases, which is completely outside the scope of semi-classical approximations.
Unfortunately, in practice, when we use semi-classical approximations, we often do not know too much about the underlying quantum theory.
For example, in QFT-CST, we do not know the exact theory of quantum gravity.
In particular, we do not always know the quantum uncertainty of a given classical quantity, for example the curvature.
As we have demonstrated in the first example, a larger quantum uncertainty means a smaller value of $\alpha_0$, which leads to a loss of unitarity sooner.
Therefore, we should really be much more careful while invoking unitarity in QFT-CST.
Our result is only the first step toward that.

We have emphasized the correlation between the uncertainty of the background, $\Delta X_B$, and the rate of losing unitarity.
Strictly speaking, our model does not allow us to tune $\Delta X_B$ independently from $m_B$.
A few other explicit examples, such as in \cite{IlgYan14} and Feynman's Lecture, have also suggested that the uncertainty in the coupling term being directly responsible for entanglement.
Thus we think our interpretation is on the right track.
Nevertheless, the relation between $m_B$ and the loss of unitarity might still give us some new insights.
Recall that in quantum field theory, the oscillator mass of a momentum mode is actually the total integrated volume.
Such quantity depends on the IR regulator, which is not only hidden from the semi-classical point of view, but also a tricky topic from the quantum perspective.
It might suggest a connection between unitarity loss and the recent developments in the IR sector of field theory, such as soft theorems and asymptotic symmetries \cite{HeLys14,HawPer16}.


\acknowledgments

We thank Robert Jefferson, Jess Riedel, Lee Smolin, Bill Unruh, and the anonymous PRD referee for discussions. 
This work is supported by the Canadian Government through the Canadian Institute for Advance Research and Industry Canada, and by Province of Ontario through the Ministry of Research and Innovation.

\bibliographystyle{utcaps}
\bibliography{all_active}

\providecommand{\href}[2]{#2}\begingroup\raggedright\begin{thebibliography}{10}

\bibitem{Haw74}
S.~W. Hawking, ``PARTICLE CREATION BY BLACK HOLES,'' {\em Commun. Math. Phys.}
  {\bf 43} (1975)  199.

\bibitem{Unr12}
W.~Unruh, ``{Decoherence without Dissipation},''
\href{http://arxiv.org/abs/1205.6750}{{\tt arXiv:1205.6750 [quant-ph]}}.

\bibitem{LidLyt93}
A.~R. Liddle and D.~H. Lyth, ``The Cold dark matter density perturbation,''
  {\em Phys. Rept.} {\bf 231} (1993)  1--105,
\href{http://arxiv.org/abs/astro-ph/9303019}{{\tt astro-ph/9303019}}.

\bibitem{Mal15}
J.~Maldacena, ``{A model with cosmological Bell inequalities},''
  \href{http://dx.doi.org/10.1002/prop.201500097}{{\em Fortsch. Phys.} {\bf 64}
  (2016)  10--23}, \href{http://arxiv.org/abs/1508.01082}{{\tt arXiv:1508.01082
  [hep-th]}}.

\bibitem{Haw76a}
S.~W. Hawking, ``BREAKDOWN OF PREDICTABILITY IN GRAVITATIONAL COLLAPSE,''
{\em Phys. Rev. D} {\bf 14} (1976)  2460--2473.

\bibitem{AMPS}
A.~Almheiri, D.~Marolf, J.~Polchinski, and J.~Sully, ``{Black Holes:
  Complementarity or Firewalls?},''
  \href{http://dx.doi.org/10.1007/JHEP02(2013)062}{{\em JHEP} {\bf 1302} (2013)
   062},
\href{http://arxiv.org/abs/1207.3123}{{\tt arXiv:1207.3123 [hep-th]}}.

\bibitem{OsuPag16}
K.~Osuga and D.~N. Page, ``{Qubit Transport Model for Unitary Black Hole
  Evaporation without Firewalls},''
\href{http://arxiv.org/abs/1607.04642}{{\tt arXiv:1607.04642 [hep-th]}}.

\bibitem{BakKod17}
D.~Baker, D.~Kodwani, U.-L. Pen, and I.-S. Yang, ``{A self-consistency check
  for unitary propagation of Hawking quanta},''
\href{http://arxiv.org/abs/1701.04811}{{\tt arXiv:1701.04811 [hep-th]}}.

\bibitem{IlgYan14}
I.~Ilgin and I.-S. Yang, ``{Energy carries Information},''
\href{http://arxiv.org/abs/1402.0878}{{\tt arXiv:1402.0878 [hep-th]}}.

\bibitem{HeLys14}
T.~He, V.~Lysov, P.~Mitra, and A.~Strominger, ``{BMS supertranslations and
  Weinberg’s soft graviton theorem},''
  \href{http://dx.doi.org/10.1007/JHEP05(2015)151}{{\em JHEP} {\bf 05} (2015)
  151},
\href{http://arxiv.org/abs/1401.7026}{{\tt arXiv:1401.7026 [hep-th]}}.

\bibitem{HawPer16}
S.~W. Hawking, M.~J. Perry, and A.~Strominger, ``{Soft Hair on Black Holes},''
  \href{http://dx.doi.org/10.1103/PhysRevLett.116.231301}{{\em Phys. Rev.
  Lett.} {\bf 116} (2016) no.~23, 231301},
\href{http://arxiv.org/abs/1601.00921}{{\tt arXiv:1601.00921 [hep-th]}}.

\end{thebibliography}\endgroup

\end{document}